# Design and verification of the HXI collimator on the ASO-S mission

Chen Dengyi [a,b], Hu Yiming [a,b], Ma Tao[a,b], Su YangYang[a,b], Jianfeng[c], Wang Jianping[d], XU Guangzhou [c], Jiang Xiankai [a,b], Guo Jianhua [a,b], Zhang Yongqiang [a], Zhang Yan [a], Chen Wei [a], Chang Jin [a], Zhang Zhe [a,b]

[a] Key Laboratory of Dark Matter and Space Astronomy,
Purple Mountain Observatory, Chinese Academy of Sciences, Nanjing 210034, China
[b] University of Science and Technology of China, Hefei 230026, China
[c] Xi'an Institute of Optics and Precision Mechanics, CAS, Xi'an 710119, China
[d] Innovation Academy for Microsatellites of CAS, Shanghai 201210, China

## ABSTRACT

A space-borne hard X-ray collimator, comprising 91 pairs of grids, has been developed for the Hard X-ray Imager (HXI). The HXI is one of the three scientific instruments onboard the first Chinese solar mission: the Advanced Space-based Solar Observatory (ASO-S). The HXI collimator (HXI-C) is a spatial modulation X-ray telescope designed to observe hard X-rays emitted by energetic electrons in solar flares. This paper presents the detailed design of the HXI-C for the qualification model that will be inherited by the flight model. Series tests on the HXI-C qualification model are reported to verify the ability of the HXI-C to survive the launch and to operate normally in on-orbit environments. Furthermore, results of the X-ray beam test for the HXI-C are presented to indirectly identify the working performance of the HXI-C.

## 1. Introduction

The Advanced Space-based Solar Observatory (ASO-S) is the first Chinese solar mission proposed for the 25th solar maximum. It is expected to be launched in early 2022 by a CZ-2D rocket at the Jiuquan Satellite Launch Center. The scientific focus of the ASO-S includes 1M2B, namely, the solar magnetic field, solar flares and coronal mass ejections. Furthermore, the ASO-S will explore the relationship among these features [1-2]. Accordingly, the ASO-S will deploy three scientific instruments: the Full-disk vector MagnetoGraph (FMG) [3], the Lyman-alpha Solar Telescope (LST) [4] and the Hard X-ray Imager (HXI).

The HXI records hard X-rays between 30 and 200 keV emitted by energetic electrons in solar flares. It has three subsystems as shown in Fig.1: a collimator (HXI-C), spectrometer (HXI-S) and electrical control box (HXI-E). The HXI-C modulates the incident X-rays with sub-collimators. The HXI-S is responsible for the count and energy observations of photons from the HXI-C. The HXI-E acquires and preliminarily processes data of the HXI on orbit [5-6].

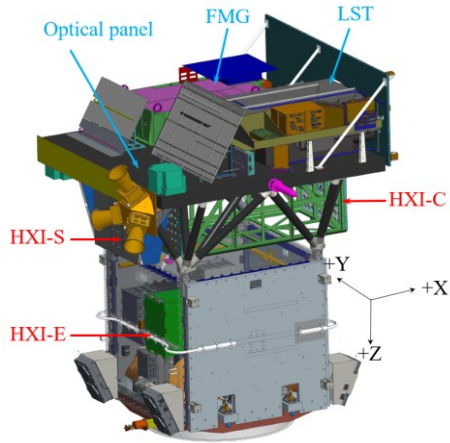

Fig. 1 Schematic view of ASO-S

The main characteristics of HXI are summarized and compared with those of its three predecessors in Table. 1. The table shows that the HXI has some advantages:

- except for the RHESSI, the HXI has the highest energy range extending to 200 keV; the range extends to 300 keV in reality;

- higher angular and temporal resolution;

- finer and more elements of grids;

Table 1. Main characteristics of HXI (comparison of similar solar missions)

|  | YOHKOH/HXT | RHESSI | Solar Orbiter/STIX | HXI/ASO-S |
|---|---|---|---|---|
| Launched time | 1991 | 2002 | 2020 | Plan 2022 |
| Type of collimator |  | Double |  |  |
| Quantity of sub-coll. | 64 | 9 | 32 | **91** |
| Pitch of grids(μm) | Finest 105μm | 34μm~2.75mm | 38μm~1mm | **36μm~1224μm** |
| Imaging method | SMC | RMC | SMC | **SMC** |
| Space resolution | 10″ | 4″~14″ | 7″ | **6″** |
| Field of view |  | Whole sun |  | **≥40′** |
| Detector | NaI(Tl) | Ge | CdTe | **LaBr$_3$** |
| Energy Range(keV) | 20~100 | 3~17000 | 4~150 | **30~200** |
| Temporal resolution | 0.5s | 2s | Up to 0.1s | **0.5s up to 0.1** |

Note: sub-coll. = sub-collimator, SMC=space modulation collimator, RMC=rotate modulation collimator;

As a core part of the HXI, the HXI-C comprises 91 grid pairs, the solar aspect system and a set of supporting structures. The biggest challenge facing the HXI-C is meeting high-precision and stability requirements. We therefore introduce the design of the instrument in detail. Moreover, as a space detector, the collimator must complete a number of tests to confirm its working performance and to identify its adaptability to the space environment. The present paper thus describes a series of tests that we conducted on the engineering qualification model (EQM) from May to August 2020.

## 2. Description of the HXI-C

### 2.1 System introduction

The Fourier transform has been widely used in high-energy solar imaging in recent decades. There are two types of such imaging, namely spatially modulating (SM) and rotating modulating (RM) methods. The SM method requires the collimator to configure quantities of grids as Fourier units while the RM method requires the collimator to cover different position angles as sufficient Fourier units through spinning. There is a contradiction between these two techniques resulting from their different principles. The ASO-S deploys three different detectors and expects a stable environment on orbit, and the HXI thus has to adopt the so-called SM technique like YOHKOH/HXT and STIX [7-8] instead of RM technique like RHESSI [9].

Subcollimators are basic components used for imaging. In this sense, more subcollimators bring about better images. However, the performance is restricted by mass and expense budgets. The HXI implements 91 subcollimators to modulate solar hard X-rays. There are 44 pairs of sin-cos subcollimators and a single set of three subcollimators. The distance between the front and rear grids is 1190 mm. The grids provide a sufficient number of u-v Fourier components for the reconstruction of images. The grids are mounted on two titanium plates at the front and rear of the framework. A Solar Aspect System (SAS) was developed owing to the low pointing accuracy of the satellite. The SAS includes a Solar Aspect (SA) and a Deformation Monitor (DM). The SAS is installed on the rear plate of the framework. With these devices, the HXI-C can record the Sun's center with the help of the SA at 2″accuracy; also, the HXI-C can obtain the relative displacement and distortion between the front and rear plates using the DM. The displacement precision reaches 2 mm and the twist precision is better than 3″. The data processing benefits much from such valuable information [10].

This paper focuses on the grid configuration and the mechanical and thermal design. After their introduction, series tests are presented. The test results are then discussed.

### 2.2 Grid configuration

The HXI-C is in fact the X-ray optics while the grid is the basic component of the HXI-C. There are several important factors affecting the grid configuration. The first is the effective detection area that determines the field of view, which needs to be as large as possible to cover the entire solar disk (around 32′). The effective detection area is also related to the spatial resolution according to the distance of 1190 mm between the front and rear grids. The second is the thickness of the grid. To achieve perfect X-ray modulation, the grid should be as thin as possible; however, this would slightly reduce the capability of X-ray blocking. This demands that the grid material be a metal with a high atomic number. Tungsten is chosen as the material for its fine mechanical and fabrication properties in this case. Ref. \cite{Su2019} conducted detailed calculation and traded off the modulation efficiency at different energies. The third is the position angles of the grids. This also affects the imaging quality. Optimization efforts were made continuously until June 2020 and detailed methods have been reported in the literature [11]. Last but not least, the distribution of the grids deserves consideration. On the one hand, the distribution should be close to the u-v distribution of the grids reported in Ref. [8]; on the other hand, effects resulting from vibration and thermal vacuum should also be taken into account. On this basis, the final grid configuration in the flight model is summarized in Table. 2 and the distribution of grids is shown in Fig. 2.

Table 2 The final grid configuration in flight model

| Pitch/μm | 36 | 52 | 76 | 108 | 156 | 224 | 344 | 524 | 800 | 1224 |
|---|---|---|---|---|---|---|---|---|---|---|
| Thickness/mm | 1.0 | 1.4 | 1.7 | 2.0 | 2.0 | 2.0 | 2.0 | 2.0 | 2.0 | 2.0 |
| Quantity/Δφ=0° | 4 | 5 | 5 | 5 | 5 | 5 | 5 | 5 | 3 | 3 |
| Quantity/Δφ=90° | 4 | 5 | 5 | 5 | 5 | 5 | 5 | 5 | 3 | 2 |
| Quantity/Δφ=120° | 0 | 0 | 0 | 0 | 0 | 0 | 0 | 0 | 0 | 1 |
| Quantity/Δφ=240° | 0 | 0 | 0 | 0 | 0 | 0 | 0 | 0 | 0 | 1 |
| Nominal spatial res./″ | 3.1 | 4.5 | 6.5 | 9.3 | 13.4 | 19.3 | 29.6 | 45.1 | 68.8 | 105.2 |
| Position angles/° | 25/70/115/160 | 5/41/77/113/149 | 32/68/104/139/176 | 23/59/95/131/167 | 14/50/86/122/158 | 5/41/77/113/149 | 23/59/95/131/167 | 5/41/77/113/149 | 23/83/143 | 53/113/173 |
| Material & Fabrication | 0.1mm tungsten foil each by laser processing in company Xi'an Micromach Technology, stacked one by one to the requirement thickness | | | | | | | | | |
| Distance | L=1190 mm (distance between the front and rear grid) | | | | | | | | | |
| Effective area | The diameter of front grid ($\varphi_f$) is set as 36 mm and for the rear grid ($\varphi_r$), it is set as 22mm. The difference between these two diameters and the distance between the front and rear grids is determined by the field of view of each sub-collimator, more than 40′. | | | | | | | | | |

(1) The definition of spatial res. is generally expressed as FWHM=p/2L;

(2) Δφ is the phase difference between the front and rear grids of a sub-collimator;

(3) Position angles is the geometry distribution of the grid assembling on rear base plate. It has been updated compared to Ref [3]. For the front grids, the angles should be the minus angle to match the rear ones, e.g. the angle of grid 36 in front should be -25/-70/-115/-160 correspondingly.

| 1224 53° | 344 59° | 224 41° | 524 41° | 344 23° | grosted glass | OPEN | 524 149° | 344 167° | 524 149° | 800 143° |
|---|---|---|---|---|---|---|---|---|---|---|
| 224 41° | 108 59° | 108 59° | 224 77° | 108 23° | 224 5° | 1224 173° | 108 167° | 108 167° | 108 131° | 224 149° |
| 108 23° | 156 50° | 36 25° | 36 25° | 76 32° | 76 32° | 156 122° | 76 139° | 76 139° | 156 122° | 344 131° |
| 800 83° | 76 68° | 156 86° | 156 86° | 524 5° | 800 23° | 344 167° | 52 113° | 52 113° | 1224 113° | 524 113° |
| grosted glass | 524 77° | 224 77° | OPEN | SA | DM | OPEN | 344 95° | 224 113° | 1224 113° | grosted glass |
| 1224 113° | 76 104° | 36 115° | 224 113° | 344 95° | 524 5° | 224 5° | 108 95° | 108 95° | 524 77° | 800 83° |
| 524 113° | 76 104° | 36 115° | 36 160° | 36 160° | 52 5° | 52 5° | 52 77° | 52 77° | 76 68° | 156 14° |
| 344 131° | 108 131° | 52 149° | 52 149° | 76 176° | 52 41° | 52 41° | 76 176° | 36 70° | 36 70° | 344 59° |
| 800 143° | 224 149° | 156 158° | 156 158° | 1224 173° | 800 23° | 344 23° | 156 14° | 524 41° | 156 50° | 1224 53° |

Fig. 2 The distribution of grid on the rear base plate

**2.3 Main support**

The greatest challenge facing the HXI-C is the high-precision assembly of grids at a distance of 1190$mm$. Error tolerance for the shift displacement should be smaller than 36 μm and that for the twist angle should be smaller than 15″ before launch following a compromise between the scientific objective and engineering actualization. For the HXI-C, these two parameters should be smaller than 20 μm and 10″ after various tests with a tradeoff made for assembly on the satellite platform. To ensure the system is sufficiently stable enough in all cases, a titanium framework was developed as the main support. Ribs were designed to enhance the rigidity and intensity of the HXI-C. Twenty thermal isolated blocks were implemented to reduce heat conduction from the satellite platform. High requirements of thermal control will be realized this way. The framework was overall cast while the mounting surface was finely finished by grinding. The flatness will be better than 0.05 mm, which will fulfill the demands of the optical platform. Grids are mounted on the front and rear base plates, which are fixed on the titanium framework with multiple screws. The base plate was fabricated from the same material (i.e. titanium) to reduce thermal expansion after threaded fastening. These materials are the so-called main support of the HXI-C. Diagrams of the main support are shown in Figs. 3 and 4.

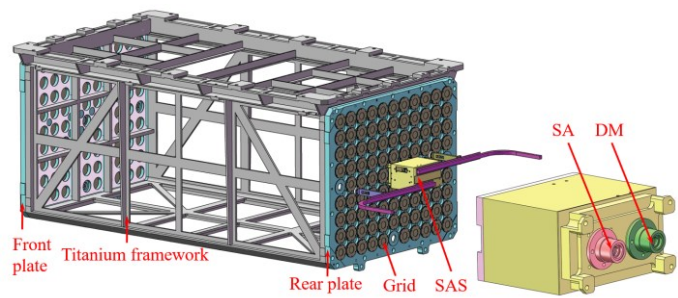

Fig. 3 (Left) Drawing of HXI-C; (Right) Titanium support main structure drawing

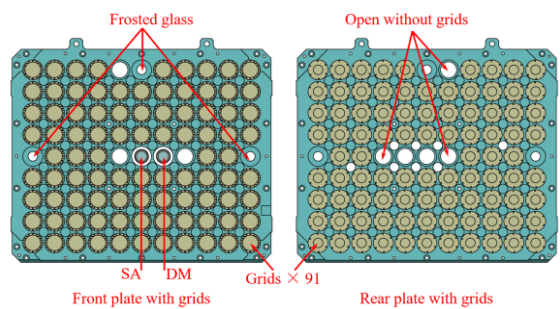

Fig. 4 Grids mounted on the front (left) and rear (right) plate. The so-called five optical units are the three frosted glass, the SA and the DM.

Finite element analysis (FEA) of the whole system was performed in 2020. The first natural frequency is about 126 Hz, which meets the requirement of the satellite payload platform that the instrument should have a first natural frequency larger than 100 Hz. The Vons Mises yield criterion is far from the yield stress for the titanium material in FEA within a dynamic environment.

**2.4 Thermal control**

The HXI-C thermal design is driven by the following considerations. (1) X-rays from the sun can be observed by the HXI-S after modulation. This requirement limits the total thickness of matter from the front grid to the rear grid. (2) Temperature gradients and differences between the front and rear should be less than 1°C. In this case, relative deformation and distortion of the grids are within the tolerance. (3) A survival temperature 5 to 35 °C is guaranteed by the satellite. The HXI-C should pass tests with property degradation within tolerance.

Taking all the factors into account, we combined active and passive thermal control method. Multiple layers were wrapped for the instrument except for the five optical units. These five units are labeled as frosted glass, the SA, and the DM in Fig. 4 and are used for measurement of the displacement and distortion of the system. In this way, the outside space and collimator are thermally isolated. Heaters are widely used and their location are given in Fig. 5. Furthermore, insulation was installed between the SAS and the framework to keep away heat from the SAS, i.e., two heat pipes were connected to irradiation panels located on the HXI-S.

On the basis of the above design, thermal analysis of the instrument was performed for the hottest and coldest cases. Figure 6 depicts the results of the FEA of the coldest case with thermal control. Results show that temperature uniformity is within 0.8 °C in expectation and allowance.

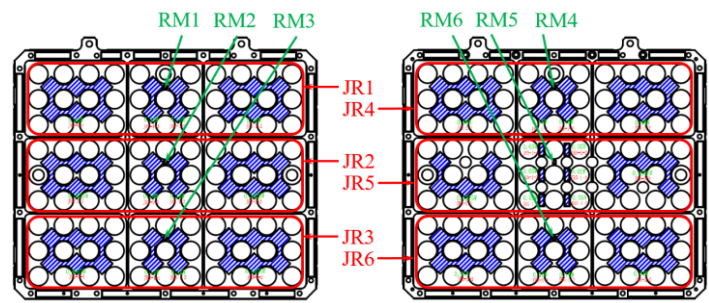

Fig. 5(a) Heating units on the front and rear plates

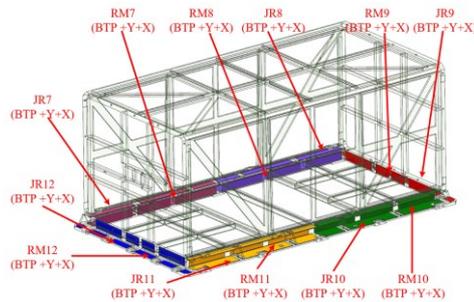

Fig. 5(b) Heating units on the framework

Note: JR *= heating area; RM *= thermistor number, BTP means bottom plate.

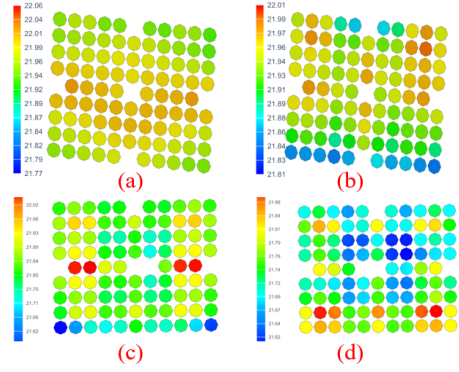

Fig. 6 Temperature distribution under cold case (up) and hot case (down). Figure (a) and (c) represent grids on front base plate; figure (b) and (d) stand for grids on rear base plate.

## 3. Tests performance

Tests on the HXI-C include environmental and X-ray beam tests. The environmental tests performed with the EQM confirm the environmental adaptability of the instrument and check the design redundancy. The X-ray beam test verifies the function of the modulation of the instrument. Triangular waves will be obtained with incident X-rays that are modulated with the grid pairs.

### 3.1 Environmental tests

To ensure that the tests are sufficiently similar, the HXI-C EQM have the same drawings, materials and integration process as the flight model. However, to reduce the fabrication time and expense, about 75\% of grids mounted on the EQM are dummy models. The models share the same material and mass distribution. In this sense, the EQM is equivalent to the flight model from the perspective of mechanical and thermal properties. Mechanical tests were first performed and then thermal tests.

3.1.1 The mechanical tests

Mechanical tests are usually vibration tests including sinusoidal and random vibrations. We adopt a model survey to investigate whether the mechanical performance of the HXI-C is reduced after these tests. The model survey also checks whether the stiffness of the instrument declines, which suggests the stabilization of the main support. Tables 3 and 4 present conditions of the mechanical test. The displacement amplitude of the vibrating table is utilized for the lowest frequency range of sinusoidal vibrations. The remaining vibrations are determined from the acceleration levels of the EQM. The sweeping rate is set as 2 oct/min.

Table 3 Sinusoidal vibration test parameters (where g stands for acceleration of gravity)

| Freq./Hz  | 5~10     | 10~14  | 14~25  | 25~100 |       |       | X direction |
|-----------|----------|--------|--------|--------|-------|-------|-------------|
| Magnitude | 15.20 mm | 6 g    | 9 g    | 3 g    |       |       |             |
| Freq./Hz  | 5~8      | 8~10   | 10~25  | 25~100 |       |       | Y direction |
| Magnitude | 23.75 mm | 6 g    | 9 g    | 3 g    |       |       |             |
| Freq./Hz  | 5~10     | 10~40  | 40~50  | 50~75  | 75~85 | 85~100 | Z direction |
| Magnitude | 10.13 mm | 4 g    | 8.5 g  | 10 g   | 6 g   | 4 g   |             |

Table 4 Random vibration test parameters

| Freq./Hz  | 20~100    | 100~600     | 600~2000  | 9.06 g RMS in each direction,          |
|-----------|-----------|-------------|-----------|----------------------------------------|
| Magnitude | +3 dB/oct | 0.1 $g^2$/Hz | -9 dB/oct | Duration time, 120s for each direction |

The mechanical tests were performed in June 2020 in Xi'an. To simulate how the instrument is fastened on the satellite optical panel, the EQM was hung upside down tightly on a fixture that was mounted on the vibrating table. A few tri-axial accelerometers were employed to measure the instrument's response and to control the test magnitude. There were four control points and 11 measuring points in total. All responses were fully recorded without exception. Little amplification was observed during the sinusoidal tests while 2.7-times enlargements were recorded during the random tests compared with the input condition. Table 5 summarizes locations of the sensors and the root-mean-square (RMS) results of the random tests. Meanwhile, the acceleration along the X direction is obviously larger than that along the Y direction and that along the Z direction. Compared with the other two directions, the X direction has the lowest rigidity because it is the principal axis. Following these tests, a three-dimensional coordinator machine surveyed the displacement and distortion of the instrument. The displacement was 12 μm and the twist was 2″ related to the absolute zero point. These values are within toleration and still have sufficient allowance.

Table 5 the acceleration response corresponding to the locations of the sensors.

| Number | Position | x direction/g | y direction/g | z direction/g |
|---|---|---|---|---|
| Ch5 | Rear plate (up 1st layer left) | 9.93 | 9.05 | 15.35 |
| Ch6 | Rear plate (up 3rd layer right) | 18.31 | 9.68 | 16.56 |
| Ch7 | Rear plate (mid 5th layer mid) | 22.44 | 12.68 | 16.02 |
| Ch8 | Rear plate (down 7th layer mid) | 24.68 | 15.74 | 16.89 |
| Ch9 | Rear plate (bottom 9th layer left) | 17.95 | 18.29 | 16.48 |
| Ch10 | Front plate(up 1st layer right) | 9.48 | 8.74 | 16.55 |
| CH11 | Front plate( up 3rd layer left) | 16.37 | 9.68 | 15.24 |
| Ch12 | Front plate( mid 5th layer mid) | 22.03 | 11.26 | 15.97 |
| Ch13 | Front plate( down 7th layer mid) | 23.49 | 13.33 | 15.42 |
| Ch14 | Front plate( bottom 9th layer mid) | 20.74 | 15.07 | 15.88 |
| Ch15 | Surface on DM | 16.68 | 19.14 | 18.27 |

3.1.2 The thermal tests

Thermal tests were carried out soon after the mechanical tests were complete. Thermal tests included thermal cycle (TC), thermal vacuum (TV), and thermal balance (TB) tests. The TC test was conducted to find material defects or rosin joints on the printed circuit board by circulating environmental stress. In contrast, the TV and TB tests were much more realistic to the space environment. Thermal design was verified in the TB test while the working performance in the toughest environments on-orbit was examined in the TV test. Figure 7 shows the profile of the TC and TV tests while Table 6 gives their test parameters.

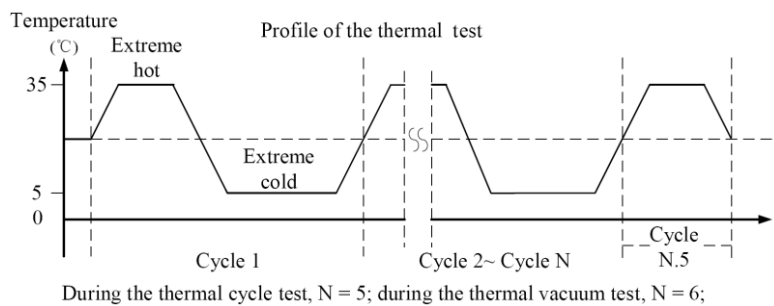

During the thermal cycle test, N = 5; during the thermal vacuum test, N = 6;

Fig. 7 Profile of the TC and TV tests. A full cycle is the period between two adjacent points with the same temperature containing an extreme hot stage and an extreme cold stage. The TC test 5.5 lasts cycles whereas the TV test includes 6.5 cycles.

Table 6 TC and TV test parameters

| Item | Pressure | Extreme Hot Temp. | Extreme Cold Temp. | Holding time of Extreme Temp. | Cycles |
|---|---|---|---|---|---|
| TC | Room Pres. | 35 ℃ | 5 ℃ | 5 h | 25.5 |
| TV | <1.3e-3 Pa | 35 ℃ | 5 ℃ | 5 h | 6.5 |

The TC and TV tests were performed in Xi'an while the TB test was performed in Shanghai. Two cases, namely the hottest case and coldest case, were considered in the TB test. Temperature results are presented in Figure 8 and 9. We conclude that the HXI-C passed the tests successfully. The temperature differences between the front and rear base plates were less than 0.8 ℃ in the hot case and less than 0.7 ℃ in the cold case. Moreover, the temperature was almost the same for the base plate and the monitored dummy models, which were made of the same material. This suggests that the requirement that the temperature differences are less than 1 ℃ was fulfilled. Relative deformation between the front and rear base plate was also measured with the three-dimensional coordinator machine. The largest displacement was 4.3μm and the relative twist was 3.8″. Both results are compared with the absolute ideal zero point. They were very small in tolerance.

In summary, the HXI-C passed all environmental tests and the instrument worked regularly during and after these tests.

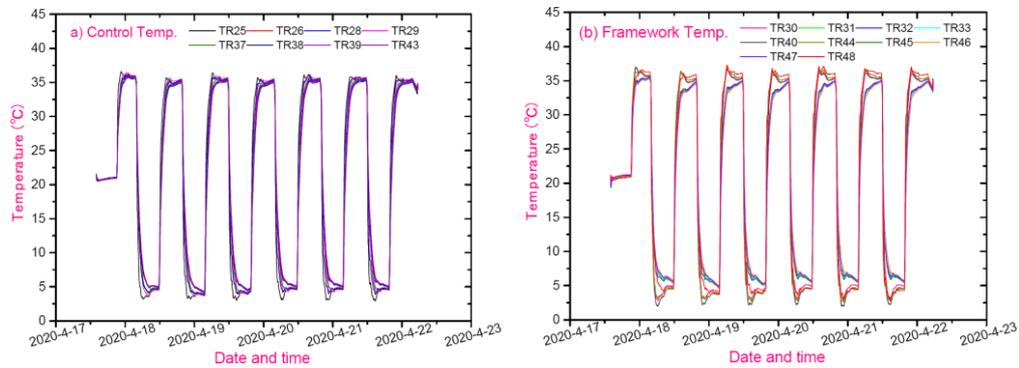

Fig. 8 Tempwerature of the framework during the TV test

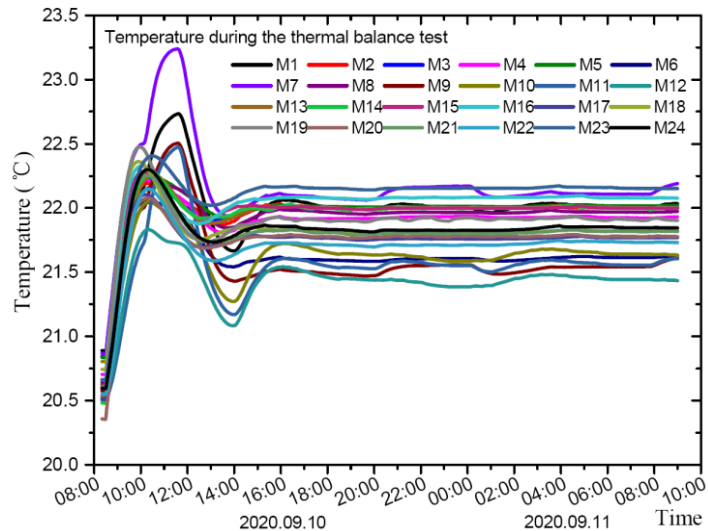

Fig. 9 Tempwerature of the front and rear base plate during the TB test. The TB test began at 08:00 on 10th, september. Cold case was carried out first until 23:59 and hot case was performed subsequently which was finished at 09:00 the next day.

**3.2 X-ray beam test**

The major concern relating to the HXI-C is the verification of the modulation function of the instrument. The X-ray beam test (BT) was proposed in this sense. A 25-meter-long stainless-steel tube with an X-ray generator was constructed for the test. In this setup, X-rays can be viewed parallel for alignment using the long tube. The test setup is shown in Figs. 10 and Fig. 11. The EQM of the HXI-C and HXI-S were employed in the test. A set of ground test instruments was developed before testing. The instruments included a high-voltage supply, data acquisition board and assistant detector module. The assistant detector module was mounted near the diaphragm on the flange of the tube to measure the counts of the generator. The diaphragm was to constrain the area of output X-rays. The main characteristics of the X-ray generator and tube are given in Table 7. The triangular waves as well as the periods were obtained with these tests. We can calculate the pitch of the tested grids from the test parameters and the periods.

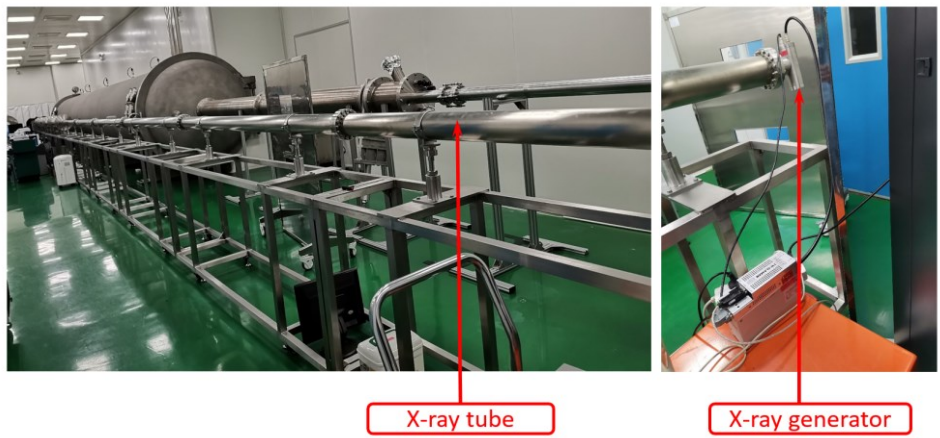

Fig. 10 Test setup of the X-ray beam test

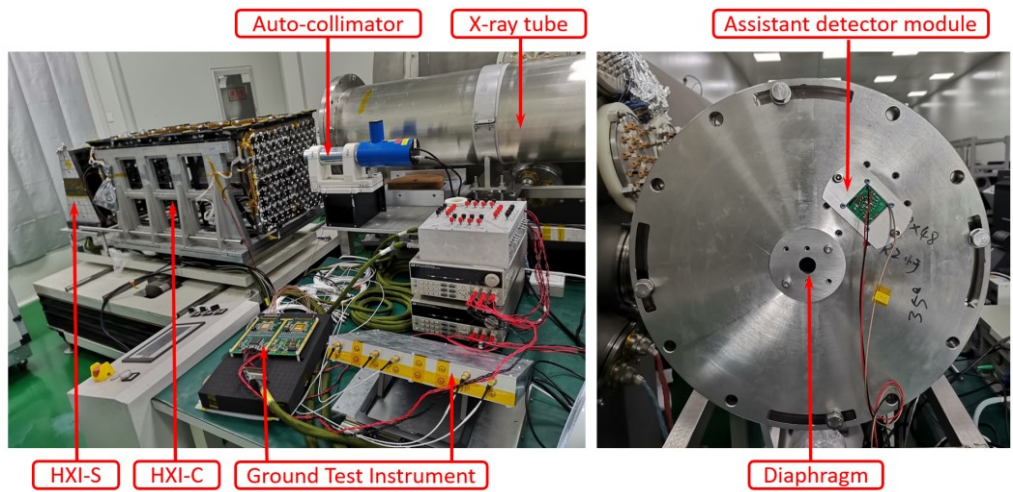

Fig. 10 Ground test instruments in the test

Table 7 Key characteristics of the X-ray BT setup

|              | Property   |                                    |
|--------------|------------|------------------------------------|
| Energy range | 10~50 keV  | Mainly 30 keV, continuously spectrum |
| Tube length  | 25m        |                                    |

| Stability | <1%@4h | about 6‰ @6h during tests |
|---|---|---|
| Flux intensity | $10^7$ cts/detector | |
| Rotating accuracy | 0.1″ / step | Resolution better than 0.05″ |

The X-ray beam test was conducted in August 2020. Seven types of grid, namely the grids p36/p52/p76/p108/p156/p224/p524[1]. were tested. Except for grids p36 and p52, we obtain all the parameters we expect following the test. Table 8 summarizes the beam test results.

Table 8 gives the X-ray beam test results. The grid type is given in the first column with the pitch. P.a. a grid's position angle. Fitting results of the grid pitch are denoted F.p. The deviation between the fitting results and the design value is denoted P.D. The amplitude of the measured triangular wave is denoted A with counts; the constant direction part is denoted B; the modulating depth is defined as 2*A/B and denoted M.D. The statistical fluctuation, denoted S.F., is calculated as Sqrt((B)/(2*A)). The diameter of the diaphragm is denoted D.d.

| Type | P.a | F.p. /μm | P.D. /μm | A | B | M.D | S.F | D.d |
|---|---|---|---|---|---|---|---|---|
| p76 | 154° | 82.31 | 6.31 | 5672.96 | 230842 | 4.92% | 4.23% | 8 mm |
| p108 | 90° | 106.34 | -1.66 | 2228.73 | 224089 | 1.99% | 10.62% | 8 mm |
| p156 | 90° | 164.08 | 8.08 | 8312.27 | 255750.48 | 6.5% | 3.04% | 10 mm |
| p224 | 90° | 223.06 | -0.94 | 25307.35 | 225450.56 | 22.45% | 0.94% | 8 mm |
| p524 | 90° | 523.80 | -0.20 | 104553.31 | 253975.93 | 82.33% | 0.24% | 8 mm |

It is clearly seen that as the pitch of grids increases, the modulating depth increases while the statistical fluctuation decreases. The result for p108 does not follow this rule because of the short testing time.

The fitting results and simulation results under the same test conditions, shown in Fig. 12/13/14, reveal that the shape of the modulating curve is well fitted. The pitch difference between the fitting results and the design value is due to the error of the rotating table. Furthermore, as the pitch increases, the modulate depth of the testing grids increases and the fitted curves become increasingly smooth. This is because of the reduced statistical error under the same intensity of the X-ray beam. The grids with pitches 224 μm and 524 μm have the best performance. This may be due to the stable X-ray beams and large periods of these grids, which suffer little from the rotation error.

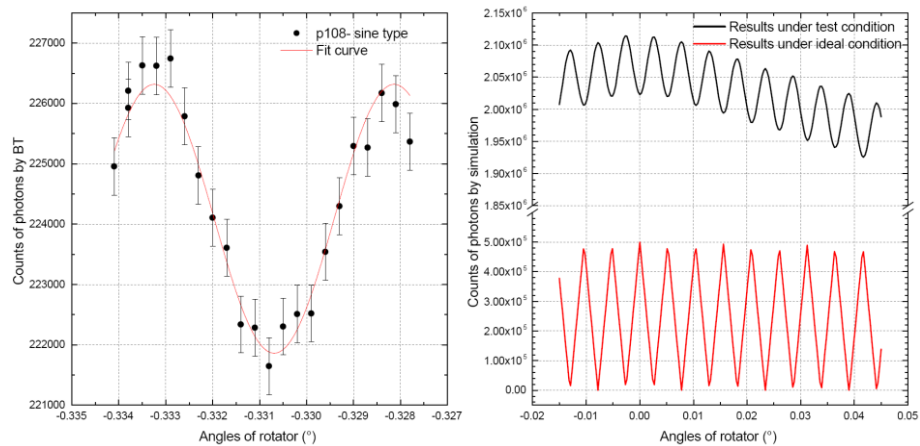

Fig. 12. The left diagram is the measured triangular wave of p108; the right diagram is the simulation curves of p108 under the same condition

---

[1] As a brief description, the grid with a pitch of, for example, 36 $\mu m$, is denoted p36.

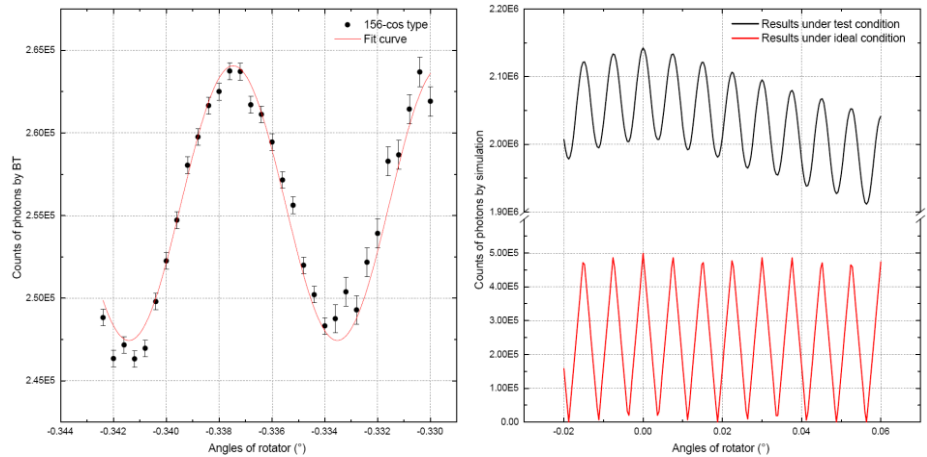

Fig. 13. The left diagram is the measured triangular wave of p156; the right diagram is the simulation curves of p156 under the same condition

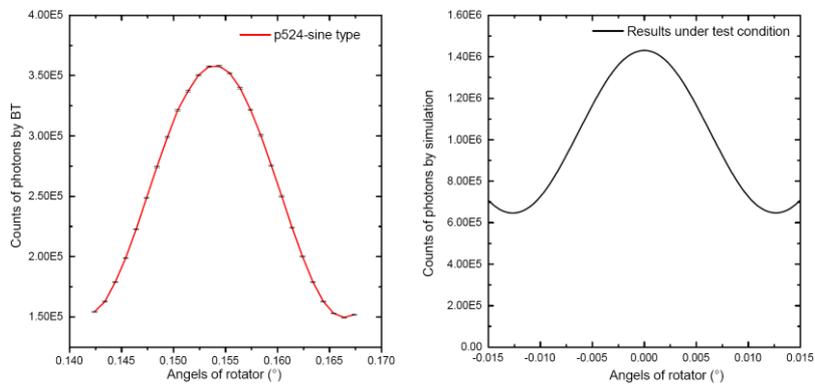

Fig. 14. The left diagram is the measured triangular wave of p524; the right diagram is the simulation curves of p524 under the same condition

However, the result is not perfect for those grids with a pitch less than 100 μm; e.g., p76 as depicted in Fig. 15 . The curve is not smooth and some points are off the cure, which may result from the poor statistics. However, the modulating profile is still distinct, and the shape and the fitted pitch are basically in accordance with the simulation results. The offset is 6.31 μm or 8.3% relative to the design value.

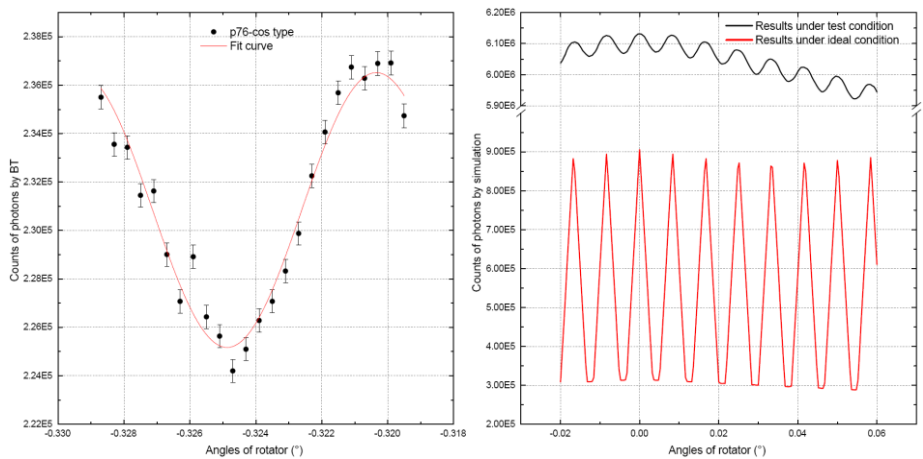

Fig. 15. Results of beam test for grid p76 (left panel: measured point and fitting curve; right panel: simulation result)

Worse still, we have not obtained any modulating profiles for p36 and p52. Multiple methods were proposed and tried with the results shown in Fig. 16/17. A qualitative evaluation is provided for p52 instead of fitted curves. Nothing can be done for the grid p36 because of its huge statistical error.

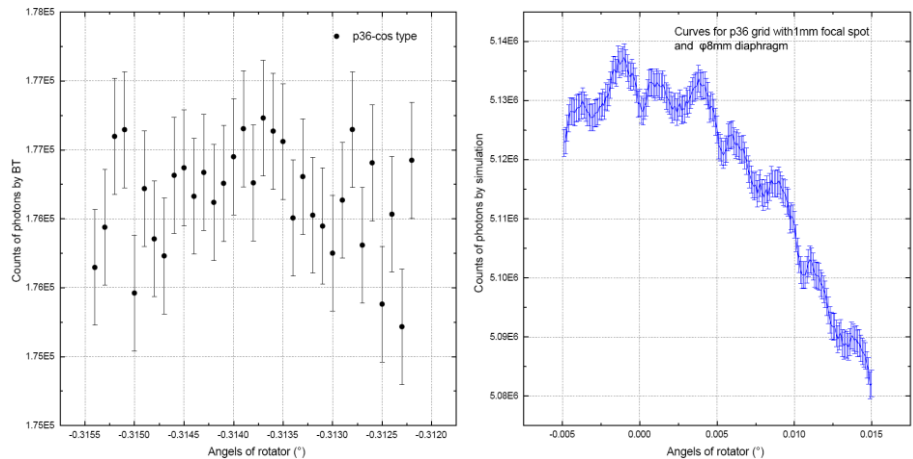

Fig. 16 The left diagram is the measured triangular wave of p36; the right diagram is the simulation curves of p36 under the same condition

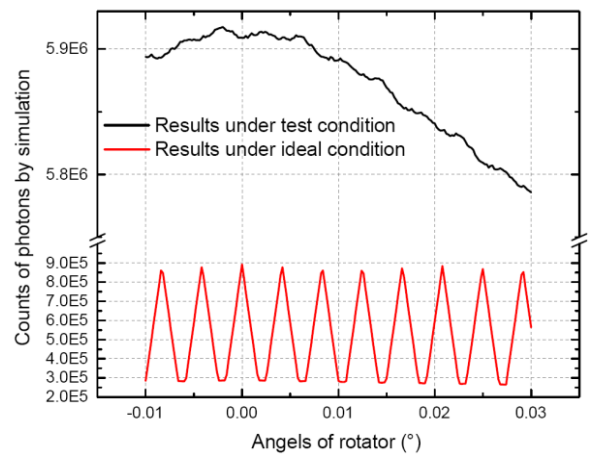

Fig. 17. Simulation results under the test condition for grid p52

The case of p36 was analyzed to search for potential reasons for the huge statistical error. There are several significant points according to the simulation results. The first point is the focal spot size and the diaphragm diameter. A proper size of the X-ray focal spot size allows the grids to be more easily measured. Figure18(a) suggests that φ0.3 and φ0.5 are more suitable than the other sizes. The size of the X-ray generator we used during the test is φ1.0 mm, which would make the recognition of X-rays difficult. Figure 18(b) shows that the diameter also affects the quality of the curve. Under the present conditions, a 5-mm-diameter diaphragm is better than a larger or smaller diaphragm for modulation. The second point is the precision of the rotating table. We test the rotating platform with the same grid clockwise and anti-clockwise. However, the results show an obvious difference in Fig. 19. The final point is the intensity of the X-ray beam. The modulation depth is so small that we have to accumulate a number of incident X-rays as shown by simulation in Fig. 20. We could not realize this accumulation because of the generator's low flux and the stability of the generator working continuously cannot be guaranteed for longer than 4 hours.

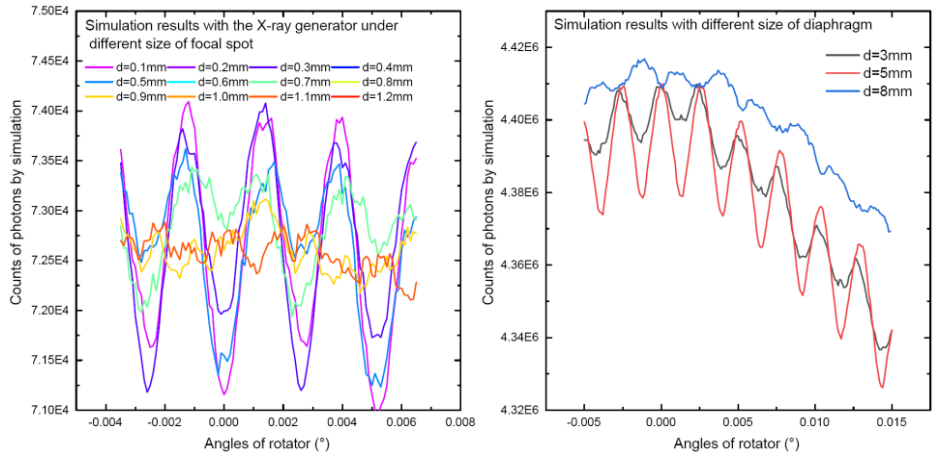

Fig. 18. Modulation curves obtained with p36 grid. The left presents results under different size of focal spot (size of diaphragm is φ3 mm); the right depicts results with different size of diaphragm (size of focal spot is φ1 mm);

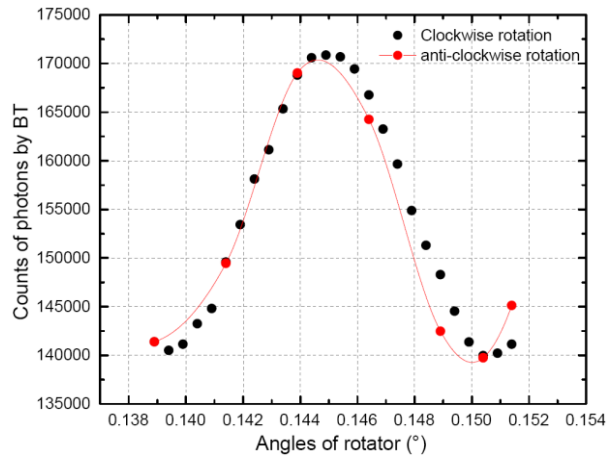

Fig. 19. Results with the grid p156. The angle decreased by clockwise (black point) and then rotated anti-clockwise with the angle increased (red point)

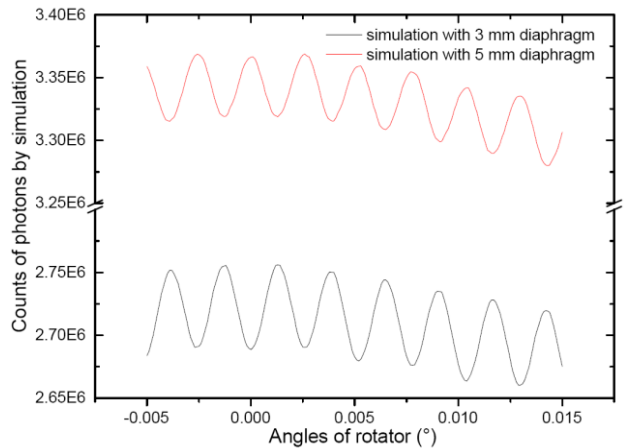

Fig. 20 Simulation results with the grid p36. Two kinds of diaphragm are set in the model. Red point is for φ3 mm and black one stands for φ5 mm. Both results require quantities of incident X-rays to reduce the effect on un-parallel beam.

A more suitable X-ray generator was purchased to solve the above problem. The focal aperture size is φ0.4 mm and the intensity of X-rays is higher by a factor of 10. Furthermore, the stable working time could is as long as 24 hours. These features will benefit testing in March 2021. We will also optimize the size of the diaphragm further before the next beam test.

In fact, we developed an auxiliary test in December 2020. The setup is much simpler than the former setup. The setup has a close alignment of subcollimator p36 with a LaBr3 detector unit. The front and rear grids are precisely positioned at a distance of 35 mm. We acquired the triangular wave as expected in Fig. 21. The pitch difference is small relative to the design value. The test confirmed that the alignment of grids is reliable and it is believed that an improved result can be obtained after the advanced X-ray generator is in place.

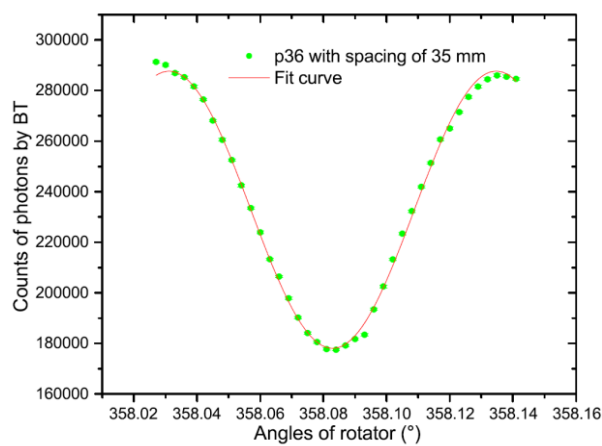

Fig. 21 Modulation curves obtained for the p36 grid at a close distance

In summary, the function of the collimator was basically verified as expected. Modulation curves for grids p36 and p52 can be obtained once the more advanced X-ray generator is introduced.

**4. Summary**

The ASO-S, the first space solar mission to be approved in China, will soon complete its phase C study and move into phase D in early 2021. As an important instrument onboard the ASO-S, the HXI-C must pass a series of tests, including environmental tests and an X-ray beam test. These tests were developed to verify the survivability and stability in tough environments and to identify the performance of the HXI-C.

This paper reports the latest design of the instrument, which has been used in the fabrication of the flight model. We also present the results of the environmental tests and X-ray beam test. The HXI-C completed all environmental tests successfully but the beam test not so well. Nevertheless, we obtained modulation curves for all grids except p36 and p52. The analysis suggests that the failure in obtaining these two modulation curves is due to the low intensity of the old generator. Different grids share the same manufacturing and alignment processes, and on the basis of the present performance, we conclude that the HXI-C will operate normally in the space environment.

### Acknowledgments

This work is supported by the Strategic Priority Research Program on Space Science, Chinese Academy of Sciences (No. XDA 15320104), National Natural Science Foundation of China (Nos. 11803093, 11973097 and 12022302) and the Scientific Instrument Developing Project of the CAS (No. 20200077). We thank Dr. Gordon J. Hurford, Yu Huang and Yufeng Shi for their useful advice on the design as well as the beam test. We also appreciate the team of beam test from　for their wonderful work.